\documentclass[12pt]{article}
\usepackage{latexsym}
\usepackage{graphics}
\usepackage{epsfig}
\usepackage{booktabs}
\usepackage{epstopdf}
\usepackage{amsmath}
\usepackage{cite}
\usepackage{hyperref}
\usepackage{amssymb}

\begin{document}

\begin{center}
\textbf{\Large 
Approximation of Invariant Solutions to the Nonlinear Filtration Equation by Modified Pad\'e Approximants
} \vspace{0.5 cm}

Sergii Skurativskyi\footnote{e-mail: \url{skurserg@gmail.com}},  Sergiy Mykulyak\footnote{e-mail: \url{mykulyak@ukr.net}},   Inna Skurativska  \footnote{e-mail: \url{inna.skurativska@gmail.com}} \vspace{0.5 cm}

S.I.Subbotin Institute of Geophysics of the National Academy of Science of Ukraine, Kyiv, Ukraine

\end{center}

\begin{quote} \textbf{Abstract.}{\small 
This paper deals with a mathematical model for oil filtration in a porous medium and its self-similar and traveling wave regimes. The model consists of the equation for conservation mass and dependencies for porosity, permeability, and oil density on pressure. 
The oil viscosity is considered to be the  experimentally expired parabolic relationship on pressure. To close the model, two types of Darcy law are used: the classic one and the dynamic one describing the relaxation processes during filtration. In the former case, self-similar solutions are studied, while in the latter case, traveling wave solutions are the focus. Using the invariant solutions,  the initial model is  reduced to the nonlinear ordinary differential equations possessing the trajectories vanishing at infinity and representing the moving liquid fronts in porous media. To approximate these solutions, we elaborate the semi-analytic procedure based on modified Pad\'e approximants. In fact, we calculate sequentially Pad\'e approximants up to 3d order for a two-point boundary value problem on the semi-infinite domain.  A good agreement of evaluated Pad\'e approximants and numerical solutions is observed. The approach  provides relatively simple quasi-rational expressions of solutions and can be easily adapted for other types of model's nonlinearity. 
}
\end{quote}

\begin{quote} \textbf{Keyword:}{\small 
 nonlinear  filtration; self-similar solution; relaxation; traveling wave; Pad\' e    approximant.
}
\end{quote}

\begin{quote} \textbf{Mathematics Subject Classification (2010):}{\small
 35K55, 35L75, 35C06, 35C07, 41A21, 70K99, 93-10.
}
\end{quote}

\vspace{0.5 cm}

\section{Introduction} \noindent  
The filtration processes are studied in many branches of science, including geophysics, biology, ecology, medicine, etc. Control of the filtration processes is at the heart of technologies applied for enhancing oil and gas recovery \cite{Barenblatt,refJPS}, cleaning polluted gas-liquid substances, and providing high-quality drugs in the biopharmaceutical industry.  Due to the significance and prevalence of these processes in nature and technological developments, theoretical studies of filtration in porous media are relevant, especially regarding the deviation of filtration flow dynamics from linear patterns. 

The complete statements of filtration problems, incorporating process nonlinearity, high intensity, and multiphasicity of liquid flows, interacting effects, and complex initial and boundary conditions, present significant challenges. This necessitates the development of new or improved tools for study.

In this research, we consider the oil filtration in a porous medium within the framework of continuous mechanics \cite{Barenblatt}, considering several nonlinear effects. The filtration model consists of the equation of motion representing the conservation of mass, the equation of state for oil, the dependencies of porosity and permeability on pressure, and finally, the Darcy law, which is considered classical or generalized. Model nonlinearity originates from the nonlinear dependence of oil viscosity on pressure which is  discovered experimentally. Note that the viscosity of reservoir oil is an important characteristic that affects the proper functioning of producing wells \cite{Danesh, Hocott}. Other fluid dynamics problems in porous media also seek to determine the influence of variable viscosity and permeability of filtrating liquids on flow behavior  \cite{Kudryashov}.

In the case of classic Darcy law,  the model in the one-dimensional case is reduced to the nonlinear filtration equation, which can be regarded as a weakly nonlinear diffusion equation \cite{Shampine,  Philip60,  Huber} 
$p_t=(k(p)p_x)_x$,  where the function $k(p)$ is the diffusion coefficient or hydraulic conductivity.
 The vast amount of studies concern boundary value problem (BVP)  on a semi-infinite domain when the model admits self-similar regimes.  
 

Another interesting class of the filtration models known  as  relaxation models  or models with memory \cite{Khuzhayorov2000, Guenoune_NDST1} has been formed when we consider the filtration processes with a relatively rapid change in parameters, the flows of non-Newtonian liquids (heavy oil, solutions of polymers, mixtures, emulsions, multiphase liquids with mass exchange between phases), and filtration in layers with a particularly complex structure (crack-porous media)  \cite{Barenblatt}. In such conditions, a delay is observed in the response of the filtration flow; in other words, there is a local nonequilibrity of the filtration process, accompanied by the relaxation of pressure and velocity. 
To incorporate process nonequilibrity, the classical Darcy law is generalized by adding the terms with the first temporal derivatives \cite{FrischRelax, SSujp2019,refJPS} describing the approach of pressure and velocity to their equilibrium values.   
As a rule, the nonequilibrium (or relaxation) filtration models do not admit  the same self-similar solutions as classical Darcy-type models. Instead, the relaxation models possess the traveling wave solutions, the structure of which is richer.  


Despite immense amount of research on the nonlinear diffusion-type equations, we rarely succeed in obtaining a general exact solution of equations, especially their hyperbolic generalizations. Therefore, there is still a need to improve existing and develop more general methods of derivation of solutions, including  the development of asymptotic approach \cite{AndrianovIntech} in combination with extensive involvement of numerical methods \cite{Jasim_NDST2}.



Thus, we aim to develop a semi-analytic approach based on the Pad\'e approximants, which were proven to be effective in many applications  \cite{Mikhlin, Boyd,AndrianovIntech}, and use it for calculating the invariant solutions describing an oil filtration with variable viscosity.


\section{The model of oil filtration and its reductions to a single equation}\label{sk:Sec1}

The mathematical model for the elastic mode of filtration   reads as follows 
\begin{equation}\label{sk:model1}
\begin{split}
(m \rho)_{t}+(\rho v)_{x}=0,\qquad \rho=\rho_{0}\left(1+C_{f}\left(p-p_{0}\right)\right), \\
m=m_{0}\left(1+C_{m}\left(p-p_{0}\right)\right),\quad
k=k_{0}\left(1+C_{k}\left(p-p_{0}\right)\right).
\end{split}
\end{equation}
Here, the system (\ref{sk:model1}) consists of  the continuity equation expressing the mass conservation law,  equation of state for a fluid, 
the dependencies of porosity and permeability on pressure.
The traditional designations are used:   $\rho$
is the fluid density, $p$ is the pressure, $v$ is the filtration velocity, $C_f$, $C_m$, and $C_k$ are the compression coefficients of the fluid, porosity, and permeability. 

To close the system,  we used the generalized Darcy law containing the description of the nonequilibrium (or relaxation) filtration processe \cite{FrischRelax,SSujp2019,refJPS,Kosterin}
\begin{equation}\label{sk:Darcy_relax}
\tau\left(v+K_\infty p_x\right)_t+v+K_0 p_{x}=0, 
\end{equation}
where the hydraulic conductivity  functions $K_0=\frac{k}{\mu}$  and $K_\infty=\theta K_0$  are related to the equilibrium and frozen diffusion coefficients, $\tau$ and $\theta$ are  constants.  In this research, we pay more attention to the oil viscosity $\mu$, assuming that it varies significantly with pressure, which prompts the consideration of nonlinear pressure dependencies for the function $\mu(p)$.

It is obvious that by dropping the relaxing terms in (\ref{sk:Darcy_relax}) out, we arrive at the classical Darcy law
\begin{equation}\label{sk:model4}
v=-K_0 p_{x}. 
\end{equation}
To simplify the problem, we reduce the model of filtration to a single equation with respect to $p$.

 Let us start with considering the model using the classic Darcy law while justifying the quadratic pressure dependence of oil viscosity $\mu$.

To specify the function $\mu$, we consider the process of oil filtration in a reservoir in the range of pressures when the oil is close to the phase transition zone which can form during depletion away from a wellbore \cite{Danesh}, p.42. We are interested in the vicinity of the phase transition point, where a single phase of oil transforms into a gas-liquid mixture. Assume that in this zone, the amount of gas phase is not enough to influence the filtration dynamics, but the oil viscosity undergoes significant changes, which are taken into account in the model.

\begin{figure}[h]
\begin{center}
\includegraphics[totalheight=4in]{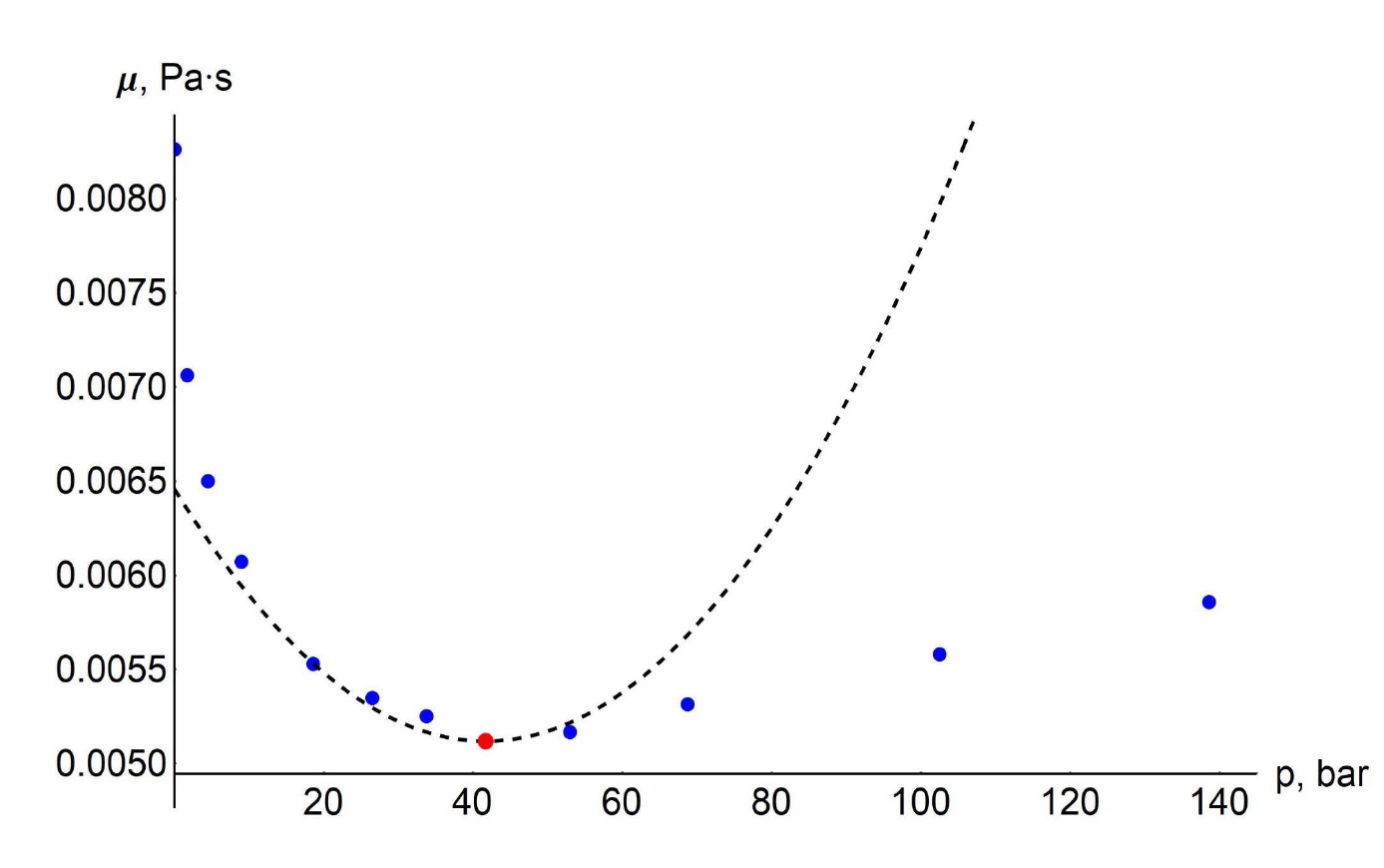}\hspace{0.5 cm} 
\end{center}
\caption{The approximation of viscosity by the parabola $\mu=\mu_{0}\left(1+a\left(p-p_{0}\right)^{2}\right)$ with the vertex $(p_0;\mu_0)=(4.16855\cdot 10^{6},0.005)$ and $a=1.507 \cdot 10^{-14}$ Pa$^{-2}$. The experiment of Hocott et al. \cite{Hocott} is  marked with filled circles, while their parabolic approximation is drawn with the dashed line.
} \label{sk:fig01}
\end{figure}

 We consider the experimental data concerning the measurements of the viscosities of oils under reservoir conditions \cite{Hocott}. Several experimental points from the paper \cite{Hocott}  (see Fig.3, curve 4) are depicted in Fig.\ref{sk:fig01}. These data confidently show the convex character of the graph of oil viscosity at pressure variations. 

In this study, we pay attention to the vicinity of the point of minimum known as a bubble point and describe the oil viscosity $\mu$ as a quadratic function of pressure. 
\begin{equation}
\label{sk:mu}
\mu=\mu_{0}\left(1+a\left(p-p_{0}\right)^{2}\right),
\end{equation}
 where $\mu_{0}$ is the viscosity at $p=p_0$, $a$ is a positive constant. 
 
To specify the function $\mu$, we approximate the experimental data in Fig.\ref{sk:fig01} by the parabola   (\ref{sk:mu}) whose vertex coincides with the minimum value of the data (Fig.\ref{sk:fig01}). 
The coordinates of the parabola vertex $(p_0;\mu_0)$  are evaluated from the experimental data quit accurately providing that  $\mu_{0}=0.005$ Pa$\cdot$s and $p_{0}=41.6855$ bar (or 4.169 MPa). The evaluation of the parameter $a$ leads us to the following value $a=1.507 \cdot 10^{-14}$ Pa$^{-2}$.

Applying the auxiliary constraints  $C_{f} C_{m} \ll 1$, the filtration model (\ref{sk:model1})  closed by 
dynamic Darcy law (\ref{sk:Darcy_relax}) are reduced to  the single partial differential equation 
\begin{equation}\label{sk:pres_relax}
\tau p_{tt}-\tau \theta [D(p) p_x]_{xt} +p_t- [D(p) p_{x}]_x=0,
\end{equation}
where 
\begin{equation*}
\begin{split}
D(p)=\kappa \frac{1+C_{k} \left(p-p_{0}\right)}{1+a\left(p-p_{0}\right)^{2}}   \mbox{  and  }  \kappa=\frac{k_0}{\mu_0 m_0 (C_f+C_m)}.
\end{split}
\end{equation*}
When $\tau=0$, i.e., it  means  that classic Darcy law (\ref{sk:model4}) is used,  it follows  from (\ref{sk:pres_relax}) the relation 
\begin{equation}\label{sk:model_red}
\begin{split}
p_{t}= \left(D(p) p_{x}\right)_{x}.
\end{split}
\end{equation}

Next, we consider invariant solutions of equations (\ref{sk:pres_relax}) and (\ref{sk:model_red}) and develop the semi-analytical procedure for their approximation using Pad\'e approximations. Let us start from the more straightforward equation (\ref{sk:model_red}) and solve BVP possessing the self-similar invariant solutions.

\section{BVP for the filtration model with the classical Darcy law and its self-similar solutions}\label{sk:Sec.2}

When equation (\ref{sk:model_red}) is subject to  
the following initial and boundary conditions 
\begin{equation}\label{sk:model_cond}
p(x, t=0)=p_{1}, \quad p(x=0, t)=p_{2}, \qquad p_{1,2}=\text{const},
\end{equation}
we arrive at the classic BVP \cite{Fujita,Lykov,Barenblatt} that admits self-similar solutions. To continue the theoretical studies, 
let us perform the substitution 
\begin{equation}\label{sk:subst1}
p(x, t)=\Omega (P+y_1), 
\end{equation}
where $\Omega=1/\sqrt{a}$ and $p_{0,1,2}=\Omega y_{0,1,2}$.

Then equation (\ref{sk:model_red}) and conditions  (\ref{sk:model_cond}) can be written in the form of   dimensionless BVP:
\begin{equation}\label{sk:BVP2}
\begin{split}
P_{t}&=\left(D(P)  P_{x}\right)_{x},  \\
P(x, t=0)&=0,\qquad  P(x=0, t)=y_{2}-y_1.
\end{split}
\end{equation}
where 
\begin{equation*}
\begin{split}
D(P)&=D(0)G(P), \qquad G(P)=\frac{1+\beta_{1} P}{1+2 \beta_{3} P+\beta_{2} P^{2}}, \qquad D(0)=\kappa \frac{1+C_{k}\Omega\left(y_{1}-y_{0}\right)}{1+\left(y_{1}-y_{0}\right)^{2}}, \\
\beta_{1}&=\frac{C_{k}\Omega}{1+C_{k}\Omega\left(y_{1}-y_{0}\right)},   \qquad 
 \beta_{2}=\frac{1}{1+\left(y_{1}-y_{0}\right)^{2}}, \qquad
\beta_{3}=\frac{\left(y_{1}-y_{0}\right)}{1+\left(y_{1}-y_{0}\right)^{2}}.
\end{split}
\end{equation*}
Further studies do not require the value of  $D(0)$ due to the special selection of solution form, while $\beta_j$ affects the solution characteristics. 

Since $a\sim 10^{-14}$ Pa$^{-2}$, then $\Omega \sim 10^{7}=10$~MPa. The values of $C_{f}$ and $C_{k}$ are of order  $10^{-10}-10^{-8}$~Pa$^{-1}$ \cite{Barenblatt}, therefore, $\beta_1$ may not be small, especially if we take into account the possibility of a negative value of $y_1-y_0$.

The remarkable feature of the model (\ref{sk:BVP2}) is that  this problem  possesses the well-known  self similar solution 
\begin{equation}\label{sk:xi}
P(x, t)=P(\xi), \quad \xi=\frac{x}{2 \sqrt{D(0) t}}, 
\end{equation}
reducing (\ref{sk:BVP2}) to the ordinary differential equation 

\begin{equation}\label{sk:red_eq}
\frac{d}{d \xi}\left(\frac{1+\beta_{1} P}{1+2 \beta_{3} P+\beta_{2} P^{2}} \frac{d P}{d \xi}\right)=-2 \xi \frac{d P}{d \xi},
\end{equation}
subjected to the conditions 
\begin{equation}\label{sk:boundary}
P(\xi=0)=y_{2}-y_{1}, \qquad P(\xi=\infty)=0.
\end{equation}

Equation (\ref{sk:red_eq})  has a long history that can be traced through the works \cite{Fujita, Lykov}. Here, we briefly remark that the construction of the solution of (\ref{sk:red_eq}) depends on the form of hydraulic conductivity function  $D(P)$. It has been known \cite{Fujita, Lykov} that the analytical representation of the solution can be obtained for the cases  $\beta_{1}=\beta_{2}=0$,  $\beta_{3}=-q, \beta_{2}=q^{2}$, and $\beta_1=0$. In other cases, equation (\ref{sk:red_eq}) is analyzed by alternative methods.

In what follows  we develop the  semi-analytic procedure for the evaluation of solutions to BVP  (\ref{sk:red_eq}) -- (\ref{sk:boundary}) utilizing the Pad\'e approximants.

\section{The Pad\'e approximant construction for the BVP self-similar solutions}\label{sk:Sec3}

To do this,  we need the integral relations representing a certain type of conservation laws for equation (\ref{sk:red_eq}).  In particular, integrating (\ref{sk:red_eq})  over the interval $(0;\infty)$, we obtain 
\begin{equation}\label{sk:ConsLaw1}
\frac{dP}{d\xi}(0)= -2\frac{1+2\beta_3 P(0)+\beta_2 P(0)^2}{1+\beta_1 P(0)}\int_0^\infty Pd\xi.
\end{equation}


In essence, the procedure is the adaptation of the approaches developed in \cite{AndrianovIntech, Mikhlin}. The direct application of the procedures outlined in the mentioned papers encounters massive symbolic calculations that do not allow for obtaining desired results or significantly exploit the peculiarities of the model.  In this research, we use the specific conservation laws and quasi-fractional Pad\'e approximants \cite{AndrianovIntech}.
 
  Thus, we are looking for the  solution of the problem    in the form of Taylor series 
\begin{equation}\label{sk:TaySer}
P=\sum_{i=0}^N r_i \xi^i, 
\end{equation}
where $P(0)=r_0$ is evaluated from the initial condition at $\xi=0$. 

Inserting it into equation (\ref{sk:red_eq}), we derive the coefficients $r_i$, $i \geqslant 2$ as functions of $r_1=P'(0)$ only. As increases $\xi$,  series (\ref{sk:TaySer}) diverges and does not describe the solution properly. Therefore, we approximate it by a Pad\'e approximant, i.e., the rational approximation for a series.  To specify the form of Pad\'e approximant, we use the additional information on the solution's behavior at infinity.

Assume that $P(\xi)$ is vanishing as $\xi$ tends to infinity. Then, from equation (\ref{sk:red_eq}), it follows that asymptotics is defined by the equation 
$\frac{d^2 P}{d \xi^2} =-2 \xi \frac{d P}{d \xi}$,
whose vanishing  solution is $P(\xi)=const \cdot \mbox{erfc}(\xi) \equiv const\cdot  \int_{\xi}^\infty \exp(-z^2) dz$. In turn, it is also valid   the asymptotics of the function  $\mbox{erfc}(\xi)\sim Q(\xi) \exp(-\xi^2)$. 

Thus, to construct the solution valid for all $\xi$, we   approximate the  Taylor series (\ref{sk:TaySer}) by the quasi-fractional Pad\'e approximant  \cite{AndrianovIntech} combining the rational Pad\'e approximant and the asymptotics $\sim \exp(-\xi^2)$
\begin{equation}\label{sk:PadeForm}
PA_{[M/M]}= \frac{\sum_{i=0}^M A_i \xi^i}{\sum_{j=0}^M B_j \xi^j} e^{-\xi^2}, 
\end{equation}
where $M$ is the order of Pad\'e approximant; $A_i$ and $B_j$ are constants. 

Let us recall that  $A_i$ and  $B_j$ depend on $r_1$ only. To evaluate $M$ coefficients of $A_i$ and  $B_j$, we need to derive $N=2M$ coefficients of the Taylor series (\ref{sk:TaySer}). Relation (\ref{sk:PadeForm}) is inserted in 
integral relations (\ref{sk:ConsLaw1}) which are solved with respect to $r_1$.  Note that the form of $PA_{[M/M]}$ can be modified further by incorporating the polynomial $\sum_{k}^L C_k \xi^k$ into the exponent of the exponential function. The constants $C_k$ can be  calculated from the auxiliary integral equations deduced from the starting equation by multiplying by $\xi^n$ and integration over $(0;\infty)$ \cite{AndrianovIntech}.

\subsection{Application of the procedure of BVP solving}

To apply the procedure developed above, we consider BVP (\ref{sk:red_eq}) -- (\ref{sk:boundary}) at $a$ and $p_0$ evaluated for the parabola of Fig.~\ref{sk:fig01}. We choose the value $p_1=2$MPa,  which lies to the left of point $p=p_0$ in Fig.~\ref{sk:fig01}.  Then we obtain 
$\Omega=1/\sqrt{a}=0.81\cdot 10^7$;  $ 
(y_1-y_0)=(20\cdot 10^5-p_0)/\Omega=-0.2677$.
To evaluate $\beta_1$, we  fix the product $C_k \Omega$ that varies in a wide range due to significant variations of  $C_k$ as mentioned above. Then, for instance, when $C_k \Omega=0.001$ then  $\beta_1=0.001$, while at $C_k\Omega=0.4$, we have  $\beta_1=0.447973$. 
Therefore, let us consider two cases. The former  when  $\beta_1$ is small enough to be  neglected and the latter when $\beta_1$ is not small. For the sake of simplicity, the initial condition $y_2-y_1$  is assumed to be 1. 
Thus, $P(0)=r_0=1$ in (\ref{sk:TaySer}) for all further studies.


%


Thus, the coefficients $r_i$ of series (\ref{sk:TaySer}) are as follows 
\begin{equation*}
\begin{split}
r_2&=r_1^2 \frac{\beta_1(\beta_2-1) +2 (\beta_2 + \beta_3)}{2(1+\beta_1)(1 +  \beta_2 + 2  \beta_3},\qquad
r_3 =-r_1 \frac{1 + \beta_2 + 2 \beta_3}{3(1 +  \beta_1)}+ \\ r_1^3& 
\frac{\beta_1^2 (3 + \beta_2^2 + 4 \beta_3) + 2 (\beta_2 + 3 \beta_2^2 + 6 \beta_2 \beta_3 + 2 \beta_3^2) + 
 4 \beta_1 (\beta_2^2-\beta_2  - 2 \beta_3 (1 + \beta_3))}{ 6 (1 + \beta_1)^2 (1 + \beta_2 + 2 \beta_3)^2},\\
%
   \dots
    \end{split}
\end{equation*}
The coefficients  $r_i$, $i \geqslant 2$ depend only on  $r_1$. However,  they quickly become cumbersome when the number $i$ increases.

Next, by relation  (\ref{sk:PadeForm}),  we construct the expression  $PA_{[M/M]}$ for the Taylor series (\ref{sk:TaySer}) in a conventional way taking into account  in addition the expansion $\exp(-\xi^2)=\sum_{n=0}^\infty (-1)^n \xi^{2n}/n!$. Let us start from the simplest case when $M=1$ and $PA_{[1/1]}=(1+A_1\xi)\exp(-\xi^2)/(1+B_1\xi)$. Then the relation for specifying $A_1$ and $B_1$ reads as follows 
\begin{equation*}
(1+r_1\xi +r_2\xi^2+\dots )(1+B_1 \xi) -(1-\xi^2+\xi^4/2+\dots)(1+A_1\xi)=0.
 \end{equation*}
Nullifying the coefficients at $\xi$ and $\xi^2$, we obtain a pair of  equations whose roots are as follows 
\begin{equation}\label{sk:caseM1}
 A_1=-\frac{1+ r_2-r_1^2}{r_1},\qquad B_1=-\frac{1 + r_2}{r_1}.
\end{equation}

\begin{figure}[tbh]
\begin{center}
\includegraphics[totalheight=1.9in]{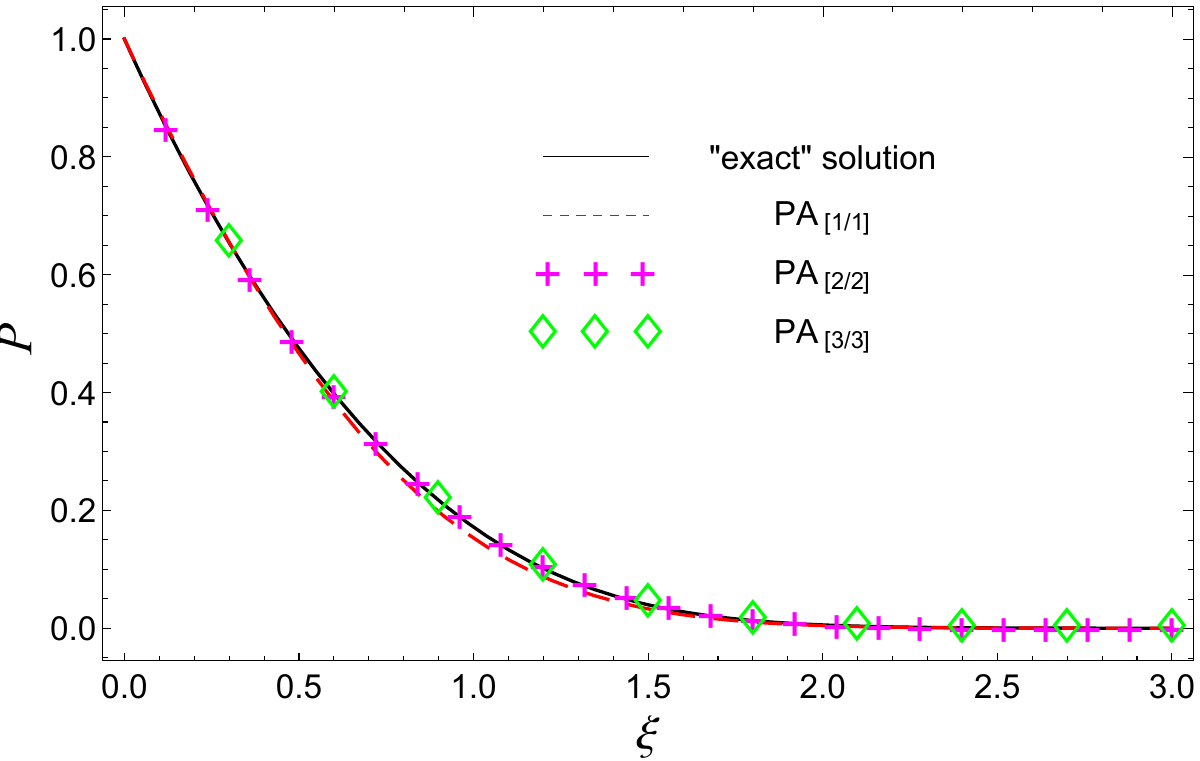}\hspace{0.5 cm}
\includegraphics[totalheight=1.9in]{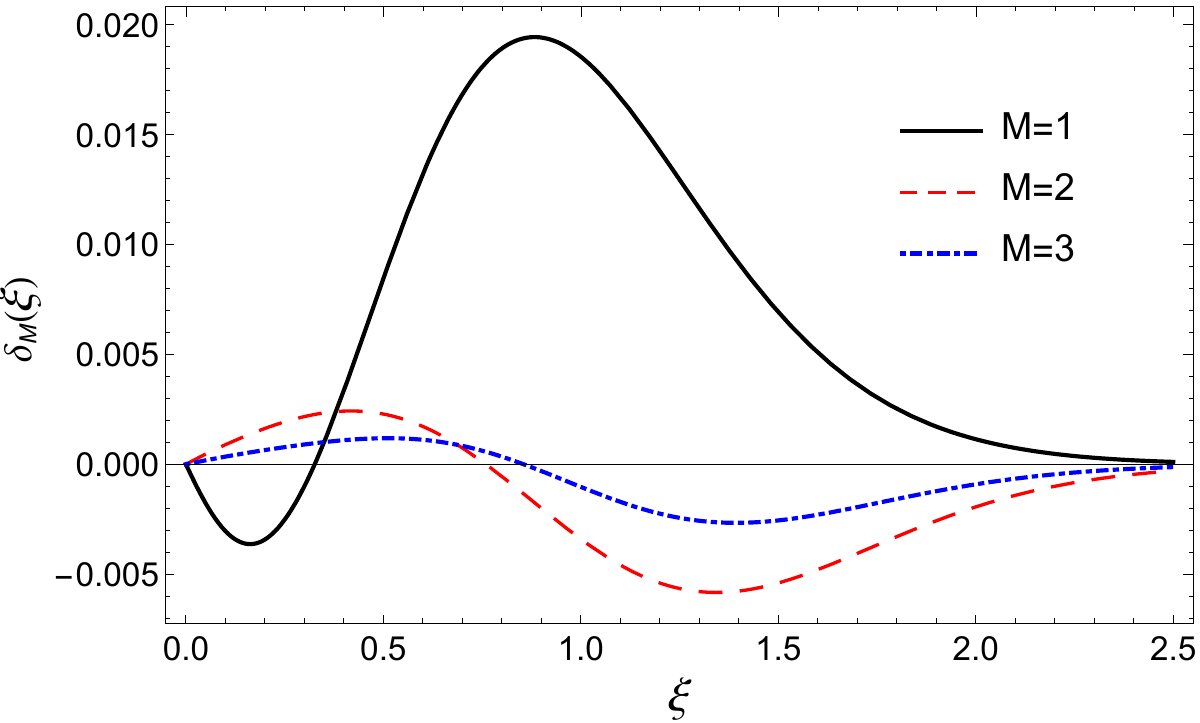}\\
(a) \hspace{8 cm} (b)
\end{center}
\caption{a: The $P(\xi)$ profiles for the solutions of BVPs (\ref{sk:red_eq}) -- (\ref{sk:boundary}) evaluated numerically  (regarded as ``exact'' solution and marked by the  solid curve) and corresponding $PA_{[M/M]}$.   b: The differences $\delta_M (\xi)=P-PA_{[M/M]}$ vs. $\xi$ for the profiles from the left panel.
} \label{sk:fig02}
\end{figure}

Let the parameter values fix as follows
\begin{equation}\label{sk:Param2}
\begin{split}
\beta_1=0.447973; \qquad \beta_2=0.933119; \qquad 2\beta_3=-0.249816.
\end{split}
\end{equation} 

Using the {\it Mathematica} commands 
\textsf{ NDSolve[Equation (\ref{sk:red_eq}), P[0]==1, P[5]==0\}, P, $\xi$,
  Method $\rightarrow $ \{"Shooting", 
   "StartingInitialConditions" $\rightarrow$ \{P[0] == 1, P$^\prime$[0] == --2\}},
we calculate the trajectory $P(\xi)$ (Fig.\ref {sk:fig02}a, solid curve) and its derivative at zero  $(r_1)_{ex}=-1.32175$, which is regarded as exact. 
 The coefficients $r_i$ of the series (\ref{sk:TaySer}) are as follows 
\begin{equation}\label{sk:Taybeta044}
\begin{split}
r_0&=1,\qquad r_1-\mbox{unknown,}\qquad 
r_2=0.3254 r_1^2, \qquad r_3= -0.3875 r_1 + 
  0.1883 r_1^3, \\
  r_4 &= -0.3783 r_1^2 + 
  0.0608 r_1^4, \dots
    \end{split}
\end{equation}

We construct $PA_{[M/M]}$ using the coefficients (\ref{sk:Taybeta044}) and insert the resulting Pad\'e approximants into the integral law (\ref{sk:ConsLaw1}). 
Solving the resulting equations with respect to $r_1$, we obtain the successive approximations for $r_1$ shown in Table \ref{sk:table2}, second row.

\begin{table}[h]
\begin{center}
\begin{tabular}{ | c |c |  c | c|  } 
  \hline
$ M$ & 1 &  2 & 3 \\ 
  \hline
  $(r_1)_M$ by  (\ref{sk:ConsLaw1}) & $-1.28324$ & $-1.33149$ & $-1.32556$  \\
  \hline
  $\eta=\left|\frac{(r_1)_{M}-(r_1)_{ex}}{(r_1)_{ex}}\right|$ & 0.0291 & 0.0073 & 0.0028
\\ 
    \hline
\end{tabular}
\caption{Values of $(r_1)_M$ and their relative errors depending on $M$ (
$(r_1)_{ex}=-1.32175$).}\label{sk:table2}
\end{center}
\end{table}
Then, using the evaluated values of $r_1$ and inserting them into (\ref{sk:Taybeta044}), we write the corresponding  Pade approximants (it is especially easy to obtain $PA_{[1/1]}$ by evaluating (\ref{sk:caseM1}))
\begin{equation}
\begin{split}
PA_{[1/1]}=\frac{1 - 0.0863411 x}{1 + 1.1969 x} e^{-x^2}, \qquad PA_{[2/2]}=\frac{1 + 0.141081 x + 0.414061 x^2}{1 + 1.47257 x + 
 0.797799 x^2}e^{-x^2},\\
 PA_{[3/3]}=\frac{1 + 0.0632479 x + 0.632533 x^2 - 0.0548591 x^3}{1 + 
 1.38881 x + 0.901649 x^2 + 0.207737 x^3}e^{-x^2}.
 \end{split}
\end{equation}

To find out the quality of the approximation, we compare the numerically derived profile $P(\xi)$   and the profiles $PA_{[M/M]}$ (Fig.~\ref{sk:fig02}a). For convenience, we depict the differences $\delta_M(\xi)=P(\xi)-PA_{[M/M]}$ in Fig.\ref{sk:fig02}b. Obviously, the deviations of the profiles from zero decrease when the order $M$ grows. In particular, at $M=3$ the difference $\delta_3(\xi)$ varies in the interval $[-0.0026; 0.0012]$, i.e.,  $P(\xi)$ and $PA_{[M/M]}$  are almost indistinguishable.

The convergence of the iteration procedure of the Pad\'e approximant evaluation  is monitored by calculating the relative error $\eta=\left|\{(r_1)_{M}-(r_1)_{ex}\}/(r_1)_{ex}\right|$, where $(r_1)_{M}=d PA_{[M/M]}/d\xi$ is the derivative of Padé approximant at $\xi=0$  and $(r_1)_{ex}=-1.32175$.
The results of the calculations are presented in Table \ref{sk:table2}, third row. Analyzing the behavior of relative errors, we see that  $\eta$ decreases when  $M$ grows. This allows one to conclude that the iteration process converges to the value $(r_1)_{ex}$. 

\section{Traveling wave solutions of the filtration model with the relaxation Darcy law and their Pad\'e approximants}\label{sk:Sec4}

Now consider equation (\ref{sk:pres_relax}), assuming that  the pressure  approaches $p_1$ at $x\rightarrow \infty$, and  transform  it using the substitution $p=\Omega  P+p_1\equiv \Omega (P+y_1)$ similar to (\ref{sk:subst1}). 
The resulting equation is as follows 
\begin{equation}\label{sk:pres_relax2}
\tau P_{tt}-\tau \theta  [D(P) P_x]_{xt}+P_t-  [D(P) P_{x}]_x=0,
\end{equation}
  where $D(P)=D(0) G(P)$ is defined in (\ref{sk:BVP2}). For our studies, we can further put  $D(0)=1$ without loss of generality.

Equation  (\ref{sk:pres_relax2}) does not admit the self-similar regimes (\ref{sk:xi}), instead, among invariant solutions, there are traveling wave regimes. 
Therefore in what follows, we consider the traveling wave solution 
\begin{equation}\label{sk:sol_relax}
p=P(\xi), \qquad \xi=x-c t,
\end{equation}
where $c$ is the phase velocity.

Inserting (\ref{sk:sol_relax}) into (\ref{sk:pres_relax2}), we get the ordinary differential equation of the third order. After integration under the condition $P\rightarrow 0$ when $\xi$ tends to infinity, we get the second order differential equation
\begin{equation}\label{sk:dyns_relax}
\tau c^2 P^\prime +c \tau \theta [G(P) P^\prime]^\prime  - c P - G(P) P^\prime=0.
\end{equation}

Thus, the problem is to approximate the forward semi-trajectory starting from $P(0)$ and approaches zero at infinity by the Pad\'e approximant. Note that such solutions can be helpful for the description of moving fronts in models for heat and mass transfer \cite{Fadai_2020}.

To develop the Pad\'e approximations for this solution, the conservation law is required. To derive it, 
we integrate  equation (\ref{sk:dyns_relax}) from 0 to infinity and arrive at the  resulting relation  
\begin{equation}\label{sk:cons_relax}
\int_0^\infty Pd\xi=\left(-\tau c^2-c \theta  G(1) P'(0)+  \int_0^1 G(x)dx\right)/c \equiv \Delta.
\end{equation}

Using the approach described in Sec. \ref{sk:Sec3},  in a vicinity of the point $\xi=0$ we look for the  Taylor series expansion $P=1+\sum_{j=1}^N r_j \xi^j$ inserting it into relation (\ref{sk:dyns_relax}). All coefficients $r_j$, $j\geq 2$ are the functions of $r_1=P'(0)$ only.  
The Pad\'e approximant now is written in the following form  
\begin{equation}\label{sk:PadeFormRelax}
PA_{[M/M]}= \frac{\sum_{i=0}^M A_i \xi^i}{\sum_{j=0}^M B_j \xi^j} e^{ H \xi}, 
\end{equation}
where the multiplier $ e^{ H \xi}$ describes the asymptotic solution's behavior as $\xi\rightarrow \infty$.
To evaluate $H$, we  linearize  equation  (\ref{sk:dyns_relax}) arriving to 
\begin{equation*}
\tau  \theta    P^{\prime \prime} + (\tau c^2-1) P^\prime - c P =0
\end{equation*}
and then the simplest  solution vanishing at infinity is $ e^{H\xi}$, where  $$H=(1- \tau c^2 -\sqrt{(\tau c^2-1) ^2+4\tau  \theta    c})/(2\tau  \theta   )<0.$$

The coefficients of the $PA_{[M/M]}$ nullify  the relation  for all $\xi$
\begin{equation*}
\sum_{j=0}^{2M} \frac{(H\xi)^j}{j!} \left(1+\sum_{j=1}^M A_j \xi^j \right) - \left(1+\sum_{j=1}^{2M} r_j \xi^j \right) \left(1+\sum_{j=1}^M B_j \xi^j \right)=0.
\end{equation*}
From this relation, the system of equations with respect to $2M$ variables $A_j$ and  $B_j$ can be extracted.  For instance, when $M=1$
we obtain 
\begin{equation}\label{sk:appr1}
A_1 - B_1 = r_1- H, \qquad   H A_1 - r_1 B_1  =r_2- H^2/2
\end{equation}
and at $M=2$
\begin{equation}\label{sk:appr2}
\begin{split}
&A_1  - B_1 =r_1- H,\qquad    A_2  - r_1 B_1 - B_2 + A_1 H=r_2  + H^2/2, \\
& - r_2 B_1 - r_1 B_2 + A_2 H + A_1 H^2/2 =r_3+ H^3/6, \\
 &- r_3 B_1 - r_2 B_2 + A_2 H^2 /2 + A_1 H^3 /6 =r_4+ H^4/24.
\end{split}
\end{equation}
Systems (\ref{sk:appr1}) and  (\ref{sk:appr2}) are linear and compatible. Thus, they possess unique solutions. 

After identifying  Pad\'e approximant (\ref{sk:PadeForm}), we insert it into the conservation law (\ref{sk:cons_relax}), where $P'(0)=r_1$, $G(1)=\frac{1+\beta_{1}}{1+2 \beta_{3}+\beta_{2}}$ = const, and  $\int_0^1 G(x)dx$ = const.

The resulting integral equation serves for the evaluation of the unknown quantity  $r_1$. It is hard to solve such an equation even numerically.  

To overcome this, we use the analytical representations for the integral term. Specifically,  for $M=1$ or $M=2$ the expression $\int_0^\infty Pdx=\int_0^\infty PA_{[M/M]} dx$ can be derived analytically. Using the {\it Mathematica} command, we obtain 
\begin{equation}\label{sk:improp_int}
\int_0^\infty PA_{[1,1]}dx=\int_0^\infty e^{H x} \frac{1+A_1 x}{1+B_1 x} dx=-\frac{A_1}{B_1 H}+\frac{A_1-B_1}{B_1^2}\mbox{ Ei }\left(\frac{H}{B_1}\right)e^{-H /B_1},
\end{equation}
where Ei$(\cdot)$ is the exponential integral function. A similar but a bit cumbersome expression can also be computed for  $M=2$. In this case, it is hard to derive the improper integral. Instead, the definite integral on the interval $[0,L]$ ($L$ is large enough, $L=4$ is used in this case) fits well.

\subsection{Pad\'e approximant construction}

Now consider the application of the approach we developed at the fixed parameters $\tau=1$, $\theta=1.5$,  $c=2.7$, and the initial condition $P(0)=1$. The parameters $\beta_{1,2,3}$ for the function $G(p)$ coincide with (\ref{sk:Param2}).

\begin{figure}[h]
\begin{center}
\includegraphics[totalheight=1.9in]{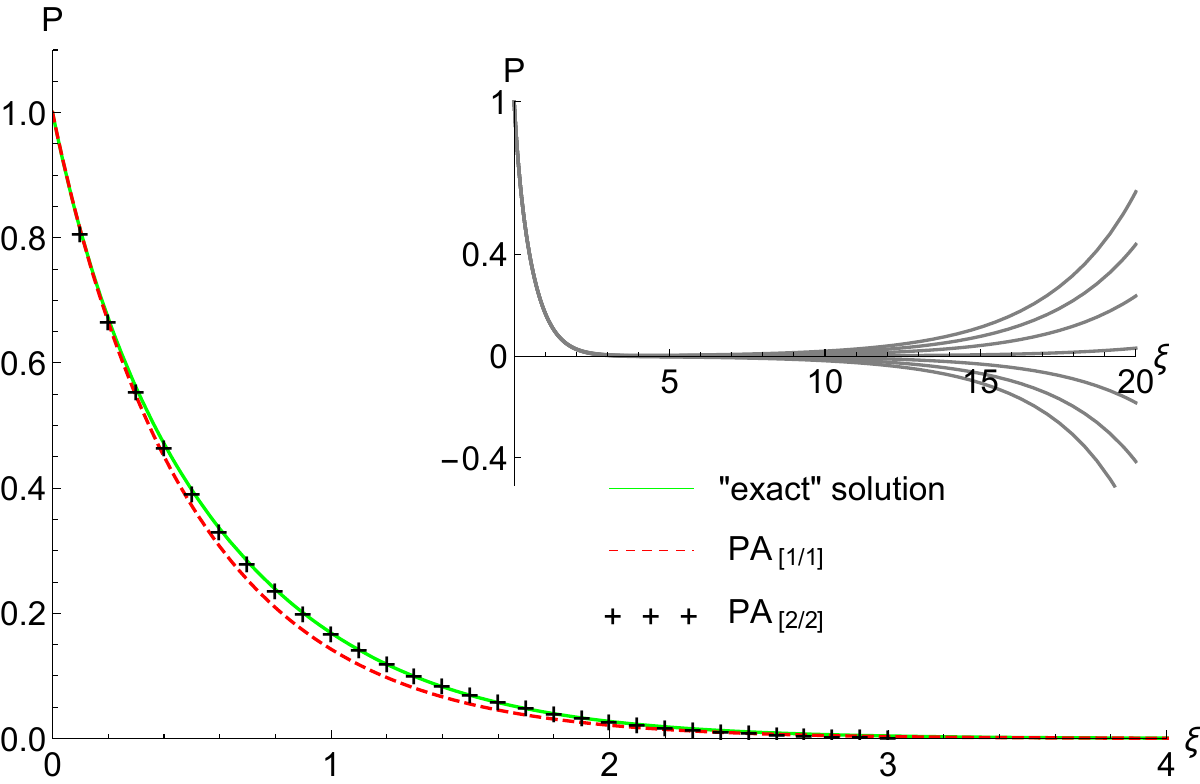}
\hspace{0.1 cm} 
\includegraphics[totalheight=1.9in]{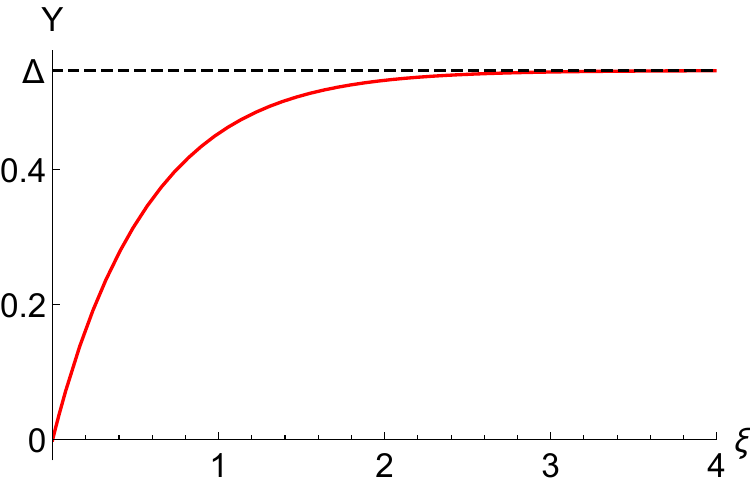}\\
(a) \hspace{8 cm} (b)
\end{center}
\caption{ (a) The profiles of the solution of equation (\ref{sk:dyns_relax}) and Pad\'e approximants   $PA_{[1/1]}$ and $PA_{[2/2]}$,  defined by (\ref{sk:PadeFormRelax}). The inset shows the bundle of trajectories starting from initial conditions $P(0)=1$ and $P'(0)$ from the range $[-2.218, -2.215]$.  (b)  The profile of $Y(\xi)$ describing the approach  of conservative quantity to its limit value $\Delta$.
} \label{sk:figRelax}
\end{figure}

To justify the existence of a trajectory vanishing at infinity, let us integrate equation (\ref{sk:dyns_relax})  at the second initial condition for the derivative $P'(0)$, which varies in the range $[-2.218, -2.215]$ with the step  0.0005. The resulting bundle of solutions, depicted in the inset of Fig.\ref{sk:figRelax}a, contains the trajectories unbounded from above and others -- from below. Then, we can conclude that a unique trajectory exists, vanishing at infinity at a certain $P'(0)$.   
To control the conservation law implementation,  we also attach the equation $dY/d\xi=P(\xi)$ with initial condition $Y(0)=0$ to equation (\ref{sk:dyns_relax}) and calculate the trajectory $Y(\xi)$ which approaches to $\Delta$ when $\xi \rightarrow \infty$ as shown in Fig. \ref{sk:PadeFormRelax}b.

 The numerically evaluated trajectory is regarded as an ``exact'' solution with which we will compare its Pad\'e approximant.
Using the {\it Mathematica} command 
\textsf{ NDSolve[$\cdot$,
 Method $\rightarrow $ \{"Shooting", 
   "StartingInitialConditions" $\rightarrow$ \{P[0] == 1, P'[0] == -1.7\}},
the trajectory we are looking for is estimated with good accuracy (Fig.\ref{sk:figRelax}a)  providing $P'(0)=-2.21658\equiv (r_1)_{ex}$.  

Finally, when the coefficients of the Taylor series $r_{1,2,3,4}$ and Pad\'e approximant $A_1$, $B_1$, and relation (\ref{sk:improp_int}) are inserted into the conservation law (\ref{sk:cons_relax}), we arrive to the equation with respect to $r_1$ possessing the root $r_1=-2.18753$. Corresponding Pad\'e approximant is as follows 
\begin{equation}
PA_{[1/1]}=\frac{0.17637 + 
   0.94988 \xi}{ 0.17637+ \xi}e^{H \xi} ,\qquad H=-1.90335.
\end{equation}
Doing in the same manner, we calculate the next $r_1=-2.22365$ and the corresponding Pad\'e approximant 
\begin{equation}
PA_{[2/2]}= \frac{1 + 2.00241 \xi + 
   1.16097 \xi^2}{1 + 2.32541 \xi + 
 0.28312 \xi^2}e^{H \xi}.
\end{equation}
Figure \ref{sk:figRelax}a exhibits the comparison of $PA_{[1/1]}$ (dashed line), $PA_{[2/2]}$ (crosses), and  ``exact'' solution (solid curve). It is obvious that $PA_{[2/2]}$ is indistinguishable from the  ``exact'' solution.

\section{Conclusion} \label{sk:Sec5}

Thus, this research considered the nonlinear model describing the oil filtration with variable viscosity in the semi-infinite domain. Model's nonlinearity was mostly determined by the quadratic pressure dependence of oil viscosity, the parameters of which  were estimated from experimental data. The model incorporating the classical Darcy law possesses self-similar invariant solutions, which allow one to reduce the starting BVP to the nonlinear BVP for an ODE on a semi-infinite domain. The traveling wave solutions are considered when the filtration model is closed by the relaxation Darcy law. The research focused on the solutions vanishing at infinity.    

We developed the semi-analytical procedure for approximating the self-similar and traveling wave solutions utilizing the modified Pad\'e approximant approach.  Comparison of the results of the procedure's application was performed with numerical   solutions. It was shown that excellent results of approximation can be achieved even under the use of low-order Pad\'e approximations (up to 3d order),  in contrast to the use of conventional rational Pad\'e approximations. Note also that the proposed procedure can be applied to the filtration equation with another form of hydraulic conductivity $D(p)$.  
Note also that the low order modified Pad\'e approximants represent relatively simple and useful expressions for the model's solutions, which are preferable to use even when the exact but cumbersome solution exists. Moreover, quasi-rational approximations are indispensable, when the further manipulations with solutions are performed. Specifically, this is important for  modeling well operation and liquid front propagation in porous media. Similar models and their solutions are also encountered in heat transfer theory \cite{Lykov,Fujita}, demonstrating the multidisciplinary nature of the research.

\section*{Acknowledgment}
The research is partly supported by the  National Academy of Sciences of Ukraine, Project 0123U100183.

\end{document}